\documentclass[11 pt]{article}

\usepackage{times}
\usepackage{epsfig}
\usepackage{graphicx}
\usepackage{amsmath}
\usepackage{amssymb}
\usepackage{mathtools}
\usepackage{caption}

\usepackage{theorem,algorithm,algorithmic}
\usepackage{amssymb,amsfonts,amsmath,latexsym}
\usepackage{graphicx}
\usepackage{mathrsfs}
\usepackage{subfigure}
\usepackage{cite}
\usepackage{fullpage}
\usepackage{tabularx}

\usepackage{placeins}
\usepackage[usenames, dvipsnames]{color}
\usepackage{xcolor}

\usepackage{verbatim}
\usepackage{hyperref}

\newcommand{\dvol} {\mbox{dvol}}
\newcommand{\la}{\langle}
\newcommand{\ra}{\rangle}

\begin{document}

\title{Comparing Three Notions of Discrete Ricci Curvature on Biological Networks}
\date{}
\author{Maryam Pouryahya, James Mathews, Allen Tannenbaum
\thanks{M.\ Pouryahya is with the Department of Applied Mathematics \& Statistics, Stony Brook University, NY; email: maryam.pouryahya@stonybrook.edu }
\thanks{J.\ Matthews is with the Department of Medical Physics, Memorial Sloan Kettering Cancer Center, NY; email: mathewj2@mskcc.org}
\thanks{A.\ Tannenbaum is with the Departments of Computer Science and Applied Mathematics \& Statistics, Stony Brook University, NY; email: allen.tannenbaum@stonybrook.edu}}

\maketitle
\thispagestyle{empty}

\section*{Abstract}

In the present work, we study the properties of biological networks by applying analogous notions of fundamental concepts in Riemannian geometry and optimal mass transport to  discrete networks described by weighted graphs.   Specifically, we employ possible generalizations of the notion of Ricci curvature on Riemannian manifold to discrete spaces in order to infer certain robustness properties of the networks of interest.
We compare three possible discrete notions of Ricci curvature (Olivier Ricci curvature, Bakry-\'Emery Ricci curvature, and Forman Ricci curvature) on some model and biological networks. While the exact relationship of each of the three definitions of curvature to one another is still not known, they do yield similar results on our biological networks of interest. These notions are initially defined on positively weighted graphs; however, Forman-Ricci curvature can also be defined on directed positively weighted networks.  We will generalize this notion of directed Forman Ricci curvature on the network to a form that also considers the signs of the directions (e. g., activator and repressor in transcription networks). We call this notion the \emph{signed-control Ricci curvature}. Given real biological networks are almost always directed and possess positive and negative controls, the notion of signed-control curvature can elucidate the network-based study of these complex networks.  Finally, we compare the results of these notions of Ricci curvature on networks to some known network measures, namely, betweenness centrality and clustering coefficients on graphs.

\section{Introduction}

In recent years, there have been tremendous efforts to elucidate the complex mechanisms of biological networks by investigating the interactions of different genetic and epigenetic factors.  Mathematical tools can help significantly with overcoming many of the challenges and can facilitate better understanding of the complexities of such networks.  This has led to the emergence of the field of network and systems biology. The mathematical methods and tools employed in networks are quite diverse and heterogeneous ranging from graph theory as abstract representations of pairwise interactions to complicated systems of partial differential equations that try to capture all details of biological interactions.

A number of notions have been employed to investigate the structure of networks such as centrality, clustering coefficient, and network connectivity.  Our study is based on several extensions of the notion of Ricci curvature on Riemannian manifolds to discrete spaces \cite{Ollivier,Kozma,Forman}. The Ricci curvature tensor of a manifold encodes the degree to which the geometry determined by a given Riemannian metric might differ from that of ordinary Euclidean space. In Riemannian geometry, sectional curvature is defined on two-dimensional tangent planes and expresses the convexity property of the distance function between geodesics.  The essential notion of Ricci curvature is the average of sectional curvatures of all tangent planes with some given direction \cite{doCarmo}.  One important aspect of Ricci curvature is that it can control the eigenvalues of the Laplace-Beltrami operator \cite{jost}.

Our use of curvature is very much involved with the theory of optimal mass transport \cite{Rachev,Villani3}.  As noted by \cite{Jordan,Otto}, the space of probability densities on a metric measure space can be endowed with a natural Riemannian metric via the Wasserstein 2-distance This is based on a seminal dynamical characterization of the Wasserstein metric given by Benamou and Brenier \cite{French} on the space of  probability densities.  Since one has a Riemannian-type metric, one can subsequently define a related notion of geodesics on the space of probability densities. As noted by Lott-Sturm-Villani \cite{LV,Strum}, if the underlying space is itself a Riemannian manifold, then via the Bochner formula, one can show that there is a fundamental relationship between changes in entropy, Ricci curvature, and the Wasserstein distance. In conjunction with the Fluctuation Theorem \cite{Demetrius}, we have considered increases of the Ricci curvature as being positive correlated with increases in system functional robustness, herein expressed as $\Delta \text{Ric} \times \Delta R \ge 0$.   An appealing feature of this correlation is that we can distinguish the cancer networks and the important nodes (targets) within such networks by a fairly straightforward computation of Ricci curvature on the networks as previously done in \cite {our_nature,Tann,NCI_60}.

There have been several efforts to define the notion of Ricci curvature that is applicable to more general metric spaces \cite{LV,Strum} \cite{Ollivier,Erbar}.   In fact, graphs are ideally suited for explaining the meaning of curvature and the appeal of this discrete feature is the fairly straightforward computation in many cases. In the present note, we employ three different approaches to define the Ricci curvature on discrete spaces, each of which is very relevant to biological networks.  Our first approach is based on a definition proposed by Olivier \cite{Ollivier} in a general framework of finite Markov processes or equivalently weighted graphs. Closely related work has been done in \cite{Lin-Yau}. The second approach follows the $\Gamma_2$-calculus of Bakry-\'Emery.  In \cite{Bakry}, Bakry and \'Emery suggested a notion analogous to curvature in the very general framework of a Markov semigroup.  Following \cite{Kozma}, these attempts lead us to define the Bakry-\'Emery Ricci curvature on graphs. The third generalization of Ricci Curvature is defined by Forman \cite{Forman} for a  large class of topological spaces called \emph{cell complexes}. Forman Ricci curvature is based on a proper interpretation of the so called Bochner-Weitzenb\"{o}ck decomposition of Laplace operator \cite{doCarmo}.  For graphs, as 1-dimensional cell complexes, we may apply this definition to study the curvature of the networks of interest \cite{Weber}.

In this work, we compare the results of all three of these notions of Ricci curvature on networks modeled as weighted graphs.   Ollivier Ricci curvature and Forman Ricci curvature are originally edge-based notions, but Bakry-\'Emery Ricci curvature is initially defined for an edge within the network.  However, the scalar Ollivier/ Forman Ricci curvature can be defined for a given node in the network.  Also, the advantage of Forman Ricci curvature is that it can also be calculated for a ``directed'' edge in the network \cite{Sreejith}. We further generalize this definition to apply the Forman Ricci curvature to the transcription network of \emph{E. Coli}.  In this network, in addition to the direction, there is a sign associated to each edge.  In other words, the direction can have positive (activator in \emph{E. Coli} network) or negative (inhibitor in the \emph{E. Coli} network) control.  We will define the \textbf{\emph{signed-control curvature}} that considers both the directionality and the signs associated to them and will calculate this notion in the \emph{E. Coli} transcription network. We regard the latter notion of curvature as one of the key contributions in the present work, that will be exploited in the future to study the properties of various networks.

In the subsequent sections, first, we review the Wasserstein distance and the Riemannian structure it induces on the space of probability densitie.  Via the Fluctuation Theorem \cite{Demetrius}, we conclude that increasing the Ricci curvature is positively correlated with increasing the robustness.  This correlation helps us to quantify the difference between robust and normal networks by computation of their Ricci curvature.
We then discuss more details of each of the three notions of the Ricci curvature on graphs.  We apply these notions on some real networks such as the cancer network generated by TCGA data and the \emph{E. Coli} transcription network.  We rank the top genes in breast carcinoma with respect to three notions Ricci curvature and compare the significant genes with these three notions. Finally, we explore the correlation between the discrete Ricci curvature on graphs and some known network measures, namely, betweenness centrality and clustering coefficient.  Interestingly, the results are similar across all three notions of Ricci curvature.

\section{Background}

In this section, we review some fundamental concepts with regards to the theory of optimal mass transport and Riemannian geometry.  We focus on an alternative formulation of Wasserstein distance by Benamou and Brenier.  This distance enables us to define a Riemannian structure on the space of probability densities due to Jordan \textit{et al.} \cite{Jordan} and Otto \cite{Otto}.  Using this structure, following the work of Lott-Sturm-Villani \cite{LV, Strum} we elucidate the relationship between the Ricci curvature and the entropy.

\subsection{Optimal mass problem and the Wasserstein distance}

The problem of optimal transport was first proposed by the civil engineer Monge in the 1780's asks for the minimal transportation cost to move a pile of soil (``d\'eblais'') to an excavation (``remblais") \cite{Evans,Rachev,Villani2,Villani3}.  Considering measure spaces $(\mathcal{X},\mu)$ and $(\mathcal{Y},\nu)$ and the cost $c: \mathcal{X} \times \mathcal{Y} \to \mathbb{R}_{+}$, Monge's problem is to find a (mass preserving) transport map $T: \mathcal{X} \to \mathcal{Y}$  that  minimizes
$$  \int_{\mathcal{X}} c(x,T(x)) \,d \mu (x). $$

Because of the difficult constraints on the mapping $T,$ Monge's problem can be non-trivial to solve and even ill-posed. In 1940's,  Kantorovich proposed a relaxation to the Monge problem that allows one to employ linear programming \cite{Evans}.  Let $\Pi(\mu , \nu)$ denote the set of all \textbf{couplings} between $\mu$ and $\nu$, that is all joint probability measures $\pi$ on $\mathcal{X} \times \mathcal{Y}$  whose marginals are $\mu$ and $\nu$.  Accordingly, the Kantorovich problem was the following linear programming problem:
$$ \inf_{\pi\in\Pi(\mu, \nu)} \int_{\mathcal{X} \times \mathcal{Y}} c(x,y) \,d \pi (x,y).$$

The cost function was originally defined in a distance form on a Polish metric space $(\mathcal{X},d)$.   This leads us to the following distance known as Wasserstein distance. For any two probability measures $\mu$, $\nu$ on $\mathcal{X}$, and $p \in [1, \infty)$, the $p$-Wasserstein distance between $\mu$ and $\nu$ is defined as
\begin {equation*}
\begin{split}
W_p (\mu ,\nu) &= \Big( \inf_{ \pi \in \Pi (\mu, \nu)} \int_{\mathcal{X}\times \mathcal{X}} d(x,y)^p d\pi(x,y) \Big) ^{1/p}
\\
&=\inf {\mathbf {E}}{\big [}d(X,Y)^{{p}}{\big ]}^{1/p} .
\end{split}
\end{equation*}
where the latter infimum is taken over all pairs of random variables $(X,Y)$ of possible couplings of $(\mu ,\nu) $.  Satisfying the three axioms of the distance function, $W_p$ defines a distance on $\mathcal {X}$ \cite{Villani3}.

The $1$-Wasserstein distance is also known as the \emph{Kantorovich-Rubinstein} distance, or \emph{Earth Mover's} distance among computer scientists. In \cite{Oll_markov}, Ollivier used the $1$-Wasserstein distance to define the (Ollivier-)Ricci curvature, which we will discuss in \ref{section:Olivier}.  The $2$-Wasserstein distance has some very remarkable properties and can be related to fluid mechanics.  In a seminal paper, Benamou and Brenier suggested an alternative numerical method to calculate the $2$-Wasserstein distance \cite{French}.  They proposed a ``dynamical" time-dependent formulation to Monge-Kantorovich Problem where one seeks for the geodesic that evolves in time between the two densities. More precisely, we assume Here $\rho(t,x)$ is the density of particles which interpolates between $\rho_0$ and $\rho_1$ and $v(t,x)$ is the velocity field at time $t$ and position $x$.
Benamou and Brenier \cite{French} formulated a dynamic computational fluid version of mass transport given by minimizing the kinetic energy of a system of particles interpolating two densities subject to a continuity constraint.  This formulation is equivalent to the Monge-Kantorovich theory for the Wasserstein 2-metric. More precisely,
\begin{equation*} \label{eq:brenier}
 \begin{split}
W_2 (\rho_0, \rho_1) = \Big( \inf_{\rho ,v} \Big{\{}\int_{\mathbb{R}^n} \int_{0}^{1} \rho(t,x) |v(t,x)|^2 dt dx:
\\
\rho (0,.) = \rho_{0} , \  \rho(1,.) = \rho_1 & \Big{\}}\Big) ^ {1/2}
 \end{split}
\end{equation*}
subject to the continuity equation:
\begin{equation} \label{eq:c1}
{\partial_t \rho} + \nabla \cdot ( \rho v) =0
 \end{equation}
where $\nabla \cdot$ stands for the divergence operator.  It turns out that the optimal conditions satisfy:
\begin{equation}  \label {Optimal-gradient}
v(t,x) = \nabla \psi (t,x),
\end{equation}
where $\psi$ is the Lagrange multiplier of the constrained optimization problem (\ref{eq:c1}).

We note that more generally, the Wasserstein distance defined above may be extended to a (complete connected)  Riemannian manifold for the space of probability densities (see our discussion below). The above distance endows the space of probability measures with a Riemannian structure as well  via the seminal work of \cite{Jordan,Otto} to which we refer the interested reader for all the details. Accordingly, one can define a notion of geodesic on the probability space.

\subsection{Ricci curvature and entropy}

Assume that $M$ is a complete connected Riemannian manifold equipped with metric $g$.
The Ricci curvature tensor provides a way of measuring the degree to which the geometry determined by a given Riemannian metric might differ from that of ordinary Euclidean space.
We refer the reader to \cite{doCarmo} for all the technical details. We just remark here that curvature controls the local behavior of the geodesics.  Generally, in the neighborhoods with positive curvature the geodesics converge, whereas when the curvature is negative they diverge.  A lower bounds on the Ricci curvature prevents geodesics from diverging too fast and geodesic balls from growing too fast in volume.  In fact, the lower Ricci curvature bounds estimate the tendency of geodesics to converge.   Interestingly, optimal transport offers a formulation of lower Ricci curvature bounds in terms of entropy that we review here.

For the measure $\mu$, the Boltzmann entropy is defined as follows:
$$ \text{Ent}(\mu)= -\int_M \rho \log \rho \; \dvol, $$
where  $ \rho=d\mu/\dvol$.  Lott, Sturm and Villani  \cite{LV, Strum} discovered a beautiful connection between Ricci curvature and the Boltzmann entropy:  Ric $\geq k $ if and only if the entropy functional is displacement $k$-concave along the $2$-Wasserstein geodesics, i.e.\ for all $\mu_0 , \mu_1 \in P_2(M)$ and $t \in [0,1]$ we have:
$$ \text{Ent}(\mu_t) \geq (1-t) \text{Ent}(\mu_0) + t \text{Ent}(\mu_1) + k \frac{t(1-t)}{2} W_2(\mu_0, \mu_1)^2, $$
where $(\mu_t)_{0\leq t \leq1}$ is the $2$-Wasserstein geodesics between $\mu_0$ and $\mu_1$. In fact, this inequality indicates the positive correlation between entropy and curvature which we express as:
$$ \Delta \text{Ent} \times \Delta \text{Ric} \geq 0. $$

\subsection{Entropy and robustness} \label{Fluc}

In this section, we review  the relation between robustness and the network entropy.  We define \emph{robustness} as the ability of a system to functionally adapt to perturbations in its environment.  As discussed by Alon \cite{Alon}, biological circuits have robust designs for which their essential function is nearly independent of biochemical parameters that tend to vary from cell to cell.  Robustness is a relative measure.  The work done in \cite{Demetrius, West} shows that in many cases the normal protein interaction networks are less robust than their cancerous analogues.

In \cite{Demetrius}, the authors proposed that robustness can be characterized by the fluctuation decay rate after random perturbations.  Their approach is as follows: Consider some perturbation in a given observable (e.g., for a graph in a node or an edge). Such changes will deviate the stationary state from its unperturbed value at time $0$.   Now let  $p_\epsilon(t)$ define the probability that the sample mean's deviation is more than $\epsilon$ from the original value at time $t$.  As t increase,  $p_\epsilon(t)$ converges to zero.  So one defines the fluctuation decay rate, noted by $R$ as follows:
$$R := \lim_{t \rightarrow \infty, \epsilon \rightarrow 0} (-\frac{1}{t} \log p_\epsilon(t)).$$
Thus, a larger value of $R$ corresponds to a smaller deviation from the steady-state condition, but this is also correlated with the larger values of entropy via the fluctuation theorem.  In fact, the fluctuation theorem indicates that the rate of decay of fluctuations to the steady state is positively correlated with entropy at the steady sate.  In other words, the fluctuation theorem implies that an increase in entropy entails a greater robustness to the perturbation of the network.  This indicates that the network entropy and robustness are positively correlated:
$$ \label{ent_rob}
 \Delta \text{Ent}  \times  \Delta R \ge 0,
$$

As we previously remarked, the work of Lott-Sturm-Villani \cite{LV,Strum} indicates the positively correlation of increases in entropy and Ricci curvature. Given the Fluctuation Theorem, one can conclude that Ricci curvature and robustness of a network are positively correlated:
\begin{equation}
\Delta R \times \Delta \text{Ric} \ge 0
\end{equation}

In conclusion, this correlation enables us to identify the network robustness and to quantify the difference between robust and normal networks by a fairly straightforward computation of Ricci curvature on the networks as was done in \cite{our_nature, Tann}.  In the following section, we study three formulations of the discrete Ricci curvature on weighted undirected graphs.

\section{Discrete versions of Ricci curvature} \label{sec:Laplacian}

In this section, we define the Ricci curvature on discrete metric measure spaces including weighted graphs.  Specifically, we assume that our network is presented by an undirected and positively weighted graph, $G=(V,E)$, where $V$ is the set of $n$ vertices (nodes) in the network and $E$ is the set of all edges (links) connecting them.
We set
\begin{eqnarray}
\label{graph-weights}
d_x &:=& \sum_z w_{xz} \nonumber \\
\mu_x(y)&:=& \frac{w_{xy}}{d_x} ,
\end{eqnarray}

The sum is taken over all neighbors $z$ of $x$, and $w_{xy}$ denotes the weight of an edge connecting $x$ and $y$ (it is taken as zero if there is no connecting edge between $x$ and $y$). Note that since the graph is undirected, we have that $w_{xy}=w_{yx}.$
Let $W_G=(w_{xy})_{1\le x,y \le n}$ be the matrix of weights, and $D_G$ be an $n \times n$ diagonal matrix whose diagonal is ($d_x$) for all $x \in E$.  Now, the  \emph{Laplacian matrix} is defined as $$L_G := D_G-W_G.$$
We define the \emph{combinatorial Laplacian matrix}, $ {\cal L}_G$, to be the negative of the Laplacian matrix.  ${\cal L}_G$ is always negative semi-definite.

\subsection{Ollivier-Ricci curvature} \label{section:Olivier}

One of the key definitions of Ricci curvature on a discrete metric measure space, is via the \textit{ Ollivier-Ricci  curvature} \cite{Oll_markov}.  As discussed by Ollivier, the Ricci curvature of a metric measure space can be defined by comparing the distance between small spheres and the distance between their centers.  He then extends this idea from the geodesic sphere to an associated probability measure at the point on a metric space $\mathcal{X}$.  Consider the  graph metric $d:V\times V \to \mathbb{R}^+$ on the set of vertices $V$ where $d(x,y)$ is the number of edges in the shortest path connecting $x$ and $y$.  For any two distinct points $x, y \in V$, the \textit {Ollivier-Ricci (OR) curvature} is defined as follows:

\begin{equation} \label{eq:Wtrans prob}
k(x,y) := 1 - \frac{W_{1}(\mu_x,\mu_y)}{d(x,y)}
\end{equation}

where $\mu_x , \,  \mu_y$ are defined in (\ref{graph-weights}).
 Here, we used the  Hungarian algorithm \cite{rubner} to compute the Earth Mover Distance on our reference networks.

Using this edge based notion of curvature, we can also define the scalar curvature of a given node in the graph as follows:
$$ S_{OR}(x):= \sum_{y} k(x,y),$$
where the sum is taken over all neighbors of $x$.

\subsection{Bakry-\'Emery Ricci curvature}

In Riemannian manifold many consequences of Ricci curvature lower bound comes from the well-known Bochner formula.  For a differentiable function $f$ on the complete Riemannian manifold $(M,g)$ we have:
\begin{equation}
 \label{Bochner}
\frac{1}{2} \Delta |\nabla f |^2 - \la \nabla f , \nabla( \Delta f) \ra = \|\text{Hess} f\|_{\text{HS}}^2 + \text{Ric}(\nabla f , \nabla f),
\end{equation}
where HS stands for Hilbert-Schmidt norm, i.e. for matrix A with its adjoint $A^{*}$, $\|A\|_{\text{HS}} = \sqrt{\text{tr}(A^{*}A)} $.
The formula has a more general version for 1-forms which is the Bochner-Weitzenb\"ock formula and it is used by Forman to define a combinatorial Ricci curvature with we will review in the following Section~\ref{sec:Forman}.  Applying the Cauchy-Schwarz's inequality, $\|\text{Hess} f\|_{HS} \geq \frac{(\Delta f)^2}{n}$, to (\ref{Bochner}) we obtain:
$$ \frac{1}{2} \Delta |\nabla f |^2 - \la \nabla f , \nabla( \Delta f) \ra \geq \frac{(\Delta f)^2}{N}+ \text{Ric}(\nabla f , \nabla f),$$
for $N \geq n$.  Therefore; by taking $N \to \infty$ we can use:
\begin{equation}\label{Bochner_ineq}
\frac{1}{2} \Delta |\nabla f |^2 - \la \nabla f , \nabla( \Delta f) \ra \geq  k |\nabla f|^2,
\end{equation}
to characterize $\text{Ric} \geq k$.

Inspired by the Bochner formula, the $\Gamma$ calculus was formulated by Bakry-\'Emery \cite{Bakry} in a way that the $\Gamma(f,f)$ is an analogue of $| \nabla f|^2$, and $\Gamma_2(f,f)$ is an analogue of $\frac{1}{2} \Delta|\nabla f|^2 - \la \nabla(\Delta f) , \nabla \ra$ in (\ref{Bochner_ineq}). Moreover, the authors of \cite{Kozma} used this $\Gamma$ calculus to give a notion of curvature on graphs.  Assume we have the graph $G=(V,E)$ and the function $f :V \rightarrow {\mathbb R}$ on the set of vertices, the discrete Laplacian $\Delta$ is defined by:
\begin{equation} \label{eq:Dis Laplacian}
\Delta f(x) = \sum_{(x,y) \in E} w_{xy}(f(y) - f(x)),
\end{equation}

For  a graph of finite number of edges and vertices, this definition coincide with the combinatorial Laplacian matrix, $\mathcal{L}_G$ that we defined before in Section \ref{sec:Laplacian}.  That is, writing $f$ as a column vector, we have:
$$ \Delta f(x) = \mathcal{L}_G f (x) $$
where $\mathcal{L}_G f (x)$ denotes the $x$th entry of the product vector $\mathcal{L}_G f$.
Now we define the two following bilinear operators for the functions $f,g :V \rightarrow {\mathbb R}$ and $x\in V$:
\begin{eqnarray}
\Gamma (f,g)(x) &:=& 1/2 [{\cal L}_G (f\cdot g)(x)-f(x) {\cal L}_G g(x)  \nonumber \\
&-& g(x) {\cal L}_G f(x)], \nonumber \\
\Gamma_2 (f,g)(x) &:=& 1/2 [{\cal L}_G (\Gamma(f,g)(x))-\Gamma(f, {\cal L}_G g)(x)  \nonumber \\
&-& \Gamma(g, {\cal L}_G f)(x)].
\end{eqnarray}

Following the Bochner formula and the inequality ($\ref{Bochner}$), we define the {\em Bakry-\'Emery-Ricci (BER) curvature} of a given node $x \in V$ as the maximum value of $k(x)$ such that:
\begin{equation} \label{eqn:discrete_ricci}
\Gamma_2(f)(x) \ge k(x) \Gamma(f)(x), \quad \forall f.
\end{equation}
where $\Gamma(f):=\Gamma(f,f)$ and $\Gamma_2(f):=\Gamma_2(f,f)$.
The later definition only makes sense if $\Gamma$ is positive semi-definite at $x$.  Also, applying $\Gamma(f): V \to {\mathbb R}$, to the operator (\ref{eq:Dis Laplacian}) we will have:
$$\Gamma(f)(x) = \sum_{y \sim x} w_{xy} (f(x)-f(y))^2.$$
Therefore, as long as our graph has positive weights we can use (\ref{eqn:discrete_ricci}) to find the BER Ricci curvature of a given node. Here, we used the convex optimization package CVX \cite{cvx} to find the BER curvature on our reference networks.

\subsection{Forman-Ricci Curvature} \label{sec:Forman}

Forman's discretization of Ricci curvature is based on the Bochner-Weitzenb\"{o}ck decomposition of the Laplace operator which is a generalization of Bochner formula to 1-forms \cite{Forman} . It is defined in general for CW-complexes. This class of spaces is broader than simplicial complexes and most importantly it includes graphs (1-dimensional CW complexes).  A $p$-cell in a CW-complex is a space that is homeomorphic to an open disk of dimension $p$. We say that two $p$-cells $\alpha_1$ and $\alpha_2$ are {\em parallel neighbors} if they only satisfy one of the following conditions:
\begin{enumerate}
\item There is a  $(p+1)$-cell $\beta$ such that $\alpha_1< \beta$ and $\alpha_2 < \beta$.
\item There is a $(p-1)$-cell $\gamma$ such that $\gamma < \alpha_1$ and $\gamma < \alpha_2$.
\end{enumerate}
where by $\alpha < \beta$ we mean that $\alpha$ is contained in the boundary of $\beta$. Forman considers the following cellular chain complex on $M$:
$$ 0 \to C_n(M, \mathbb{R}) \xrightarrow{\partial} C_{n-1}(M, \mathbb{R}) \xrightarrow{\partial} \ldots \xrightarrow{\partial}  C_0(M, \mathbb{R}) \to 0$$
Replacing the derivatives by the boundary operator and the $p$-forms by $p$-cells, Forman defines an analogous notion of \textit{ Laplace-Beltrami } operator on $C^p (M)$ as follows:
$$ \square_{p} = \partial \partial^* + \partial^* \partial: C^p (M) \to C^p (M).$$
where $\partial^*$ is the formal adjoint operator of $\partial$.

The Bochner-Weitzenb\"ock theorem gives the following decomposition of Laplace-Beltrami operator on $p$-forms \cite{jost}:
$$ \Delta_{p} = \nabla^*_{p} \nabla _{p} + F_p $$
where $\nabla_{p}$ is the covariant derivative operator and $F_p$ is a linear operator involving only curvature.  Specially, for a $1$-form, $\omega \in T_{x}^*(M)$, we have:
\begin{equation} \label{eq:1form}
 \mathcal{F}_1(\omega) = \la F_1(x)\omega,  \omega \ra = \text{Ric} (\omega).
 \end{equation}
In combinatorial settings, Forman \cite{Forman} shows that there is a canonical decomposition:
$$ \square_{p} = B_{p} + F_{p}, $$
where $B_p$ is a nonnegative operator.  In analogy with Bochner-Wetzenb\"rock formula, for any $p$-cell $\alpha$, he defines:
$$ \mathcal{F}_p(\alpha) = \la F_p(\alpha),  \alpha \ra.  $$ He finally reduces this definition to an explicit calculation of $\mathcal{F}_p$ as follows:
\begin{align*}
{\cal F}_p(\alpha) := \#\{(p+1)\mbox{-cells } \beta> \alpha \} +  \#\{(p-1)\mbox{-cells }
\gamma<\alpha \} \nonumber \\
- \#\{\mbox{parallel neighbors of $\alpha$}\}.
\end{align*}

Similar to formula (\ref{eq:1form}), for the special case of $p=1$, Forman defines the {\it Ricci curvature} of each edge $e$ (1-cell) by
$$
\mbox{Ric}(e) = {\cal F}_1(e)
$$
Therefore, we have:
$$ \text{Ric}(e)= \# \{ 2\text{-cells} \,  f > e\} +2 - \# \{ \text{parallel neighbors of} \, e\}. $$
Forman also extends this definition to the weighted case where each cell has been assigned a positive weight.  We can use this formula to define the curvature of a network \cite{Weber, Sreejith}, .  Assume that as before our network is presented by a weighted graph $G=(V,E)$.  The {\it Forman-Ricci (FR) Curvature} of the graph $G$ is defined as follows:
\begin{equation}
\label{Forman-curvature-weighted}
\begin{split}
\text{Ric}(x,y) &= w_{xy} \Bigg( \frac{w_x}{w_{xy}} - \sum_{z \neq y} \:  \frac{w_x} {\sqrt{w_{xy} w_{xz}}}  \Bigg)
\\
& + w_{xy} \Bigg( \frac{w_{y}}{w_{xy}} - \sum_{s \neq y}  \: \frac{w_{y}} {\sqrt{w_{xy} w_{sy}}}  \Bigg),
\end{split}
\end{equation}
where the sum is taken over all neighbors $z$ of $x$ and $s$ of $y$ except $y$ and $x$, respectively. $w_{x}$ denotes the weights associated with the node $x$, which is calculated as the sum of the weights of the incident edges ($d_x$ in (\ref{graph-weights})) divided by the degree of the node.  $w_y$ is defined similarly.

Furthermore, similar to OR curvature one can define the scalar FR curvature for a given node $x$ in the network as follows:
\begin{equation}
\label{scalar_FR}
S_{FR}(x):= \sum_{y} {\text {Ric}}(x,y),
\end{equation}
where, again, the sum is taken over all neighbors of $x$.


\section{Empirical studies}

In this section, we apply the Ricci curvature analysis to some model and real networks.  We investigate seven networks composed of cancer specific genes provided by Memorial Sloan Kettering Cancer Center \cite{sander} which have been a subject of investigation in previous work \cite{our_nature, Tann, MTNS}.  The data initially consists of $1600$ genes and RNA-Seq  data from 3000 samples of primary tumor and adjacent normal sample from seven distinct tissues. We chose $500$ cancer-related genes of this data from COSMIC cancer gene mutation data \cite{Cosmic}. The TCGA gene expression  (The cancer Genome Atlas, \href{https://cancergenome.nih.gov/}{https://cancergenome.nih.gov/}) is derived for seven corresponding tumor types: breast invasive carcinoma [BRCA], head and neck squamous cell carcinoma [HNSC], kidney renal papillary cell carcinoma [KIRP], liver hepatocellular carcinoma [LIHC], lung adenocarcinoma [LUAD], prostate adenocarcinoma [PRAD], and thyroid carcinoma [THCA].  The network is constructed by (Spearman) correlation values of gene-to-gene expressions in cancerous and normal tissues across all samples within a given genotype. We further used the transformation of $\frac{(1+corr(i,j))}{2}$ for the genes $i$ and $j$ to make a positively weighted network. The underlying biological gene-to-gene interactions are provided by the Pathway Commons Project (\href{http://www.pathwaycommons.org/pc2/downloads}{http://www.pathwaycommons.org/pc2/downloads}).
We downloaded the binary relationships between pairs of genes that are included in our cancer gene set. Here, we show that we get consistent results applying the OR, BER and FR curvatures on these tumor and normal networks. The OR results are directly taken from \cite{our_nature,Tann}.

As we will discuss in more detail in the upcoming section, FR curvature can also be applied to directed networks.  We further generalized this notion to the signed-control network which have positive/ negative directions.  We apply this notion of curvature on the transcription network of Escherichia coli ({\it E. coli} derived from the Uri Alon laboratory website at the Weizmann Institute \cite{E.Coli_Alon}.  In this network the `nodes' are operons which contain one or more structural genes transcribed on the same mRNA. Each `edge' is directed from an operon that encodes a transcription factor to an operon that it directly regulates \cite{Alon2}. This network consists of 423 nodes (operons) and 578 edges. In addition to the direction of the regulation, the regulator could be an activator (promotor) or repressor (inhibitor).  We generalize the FR curvature to the one that also considers the signs of these directions.  As discussed by Uri Alon \cite{Alon} the transcription network of E. Coli as a biological network exhibits a great degree of robustness.

Finally, we compare the notion of Ricci curvature on networks to two common measurement on networks, betweenness centrality and clustering coefficient.  To this end, we consider the scale-free network of Barab\'asi-Albert (BA) model \cite{BA} and the network of breast cancer mentioned above.  The  BA model is an algorithm for generating a random scale-free network, in which the degree distribution (for large values of $k$) is a power law of the form:
$$ P(k) \, \sim \, k^{-3} $$
where $P(k)$ is the fraction of nodes in the network having $k$ connections to other nodes. The network begins with an initial connected network. New nodes are connected to existing nodes at a time with a probability that is proportional to the number of links that the existing nodes already have. Consequently, the new nodes prefer to attach themselves to the already heavily linked nodes known as hubs. We discuss the results of these investigations in the following sections.

\subsection{Application to cancer networks}

We investigate the three notions of Ricci curvature on the seven different genotype networks we discussed earlier.  Table~\ref{table:6} provides the results of the average deference between cancer and normal network (cancer-normal) in terms of the Ricci curvature.  It has been posited in several studies \cite{Demetrius, West}  that cancer networks exhibit a higher degree of robustness than their analogous normal tissue network.  Here, we investigate this hypothesis by comparing the three discrete curvatures on cancer and normal networks of the TCGA data.  As it is shown in Table~\ref{table:6}, all three Ricci curvatures (OR, BER and FR curvature) have higher values in the seven cancer networks compared to the normal ones.

\begin{table}
\centering
 \begin{tabular}{ |p{3cm}||p{1cm}|p{1cm}|p{1cm}|}
 \hline
 \bf{ {Cancer Type}} & \bf {$\Delta$ Average OR Curvature} & \bf {$\Delta$ Average BER Curvature} & \bf {$\Delta$ Average FR Curvature}\\ [0.5ex] 
 \hline\hline
 Breast Carcinoma & 0.012  & 0.182  & 13.022 \\
 \hline
Head/Neck Carcinoma &  0.004  & 0.116  & 9.100  \\
 \hline
Kidney Carcinoma & 0.010 & 0.217 &  7.711  \\
 \hline
 Liver Carcinoma &  0.008 & 0.227  & 3.136 \\
 \hline
Lung Adenocarcinoma & 0.013 & 0.320  & 7.898 \\
\hline
Prostate Adenocarcinoma & 0.009 & 0.179  & 7.368 \\
\hline
 Thyroid Carcinoma & 0.006 & 0.133  & 2.969 \\ [1ex]
 \hline
\end{tabular}
\caption{All seven cancer networks have a higher average Ricci Curvature than the complementary normal networks.}
\label{table:6}
\end{table}

\subsubsection*{Top ranked genes in breast carcinoma }

We also ranked the top genes in breast carcinoma with respect to three notions of Ricci curvature.  These genes has been ranked based on their curvature's difference in cancer and normal networks (cancer- normal).  In fact, these are the nodes expected to have the most contribution in the robustness of the cancer network compared to the normal network.  Since the OR and FR curvature are initially edge based, we provide the top pair in Tables \ref{table:8} and \ref{table:7}.  Table~\ref{table:9} provides the top node-based BER curvature.
There are some similarities between the top ranked genes with respect to FR curvature and BER curvature, namely, ALDH2, NDRG, CLTCL1, KIF5B, PPARG, PTPN11, JAK1, PIK3CA, SDHB, EPS15, ERG, and HIP. There are some similarities between the top ranked genes with respect to FR curvature and OR curvature, namely IDH1, RUNX1, HMGA1, SDHB, EPS15, and ERG.
There are three genes, SDHB, EPS15, and ERG found among the top ranked genes with respect to all three  FR, BER and OR curvatures.

A number of these genes have known clinical implications with regards to breast cancer.  For example, the EPS15 gene encodes for the Epidermal growth factor receptor subset 15  (EPS15 protein) \cite{EPS15}. EPS15 plays a crucial role in the degradation of growth factor receptors. Furthermore, the development of a wide variety of malignancies, including breast cancer, is known to be associated with the over-expression of receptor tyrosine kinases (RTKs), which are largely comprised of growth factor receptors. Furthermore, the prognostic value of EPS15 has been reported given the notion that over-expression of EPS15 is significantly associated with a favorable clinical outcome with regards to breast cancer.
Also, mutations of the Succinate Dehydrogenase B (SDHB) gene lead to a reduction in the amount of SDHB protein in the cell and loss of SDH enzyme activity. Within mitochondria, the SDH enzyme plays an important role in energy metabolism, linking the the Krebs cycle with oxidative phosphorylation. Loss of function of this gene and subsequent decreased SDH enzyme activity results in abnormal hypoxia signaling and increased formation of tumors. Furthermore, the SDHB gene is a noted tumor suppressor gene and has been identified as a culprit in various cancers including Gastrointestinal Stromal Tumors (GIST), malignant pheochromocytoma and paraganglionomas, as well as renal cell carcinoma. There are case reports in the literature of breast cancer involving SDHB mutations, although further studies in this regard are warranted \cite{Kim, SDHB}. Of note, one could provide this comparison by the node-based scalar OR and FR curvature as we discussed before. However, doing so seems to miss some of the original information;  therefore, we prefer to find the top genes with regards to the original results.   Also, there are some important cancer-related gene mutations known to play a significant role in breast carcinoma such as BRCA1 and BRCA2 which are not ranked among the top ranked genes with regards to curvature.  This potentially indicates the limitation of node-wise study of the curvature on the networks compared to the global one. This could be due to the fact that the node-wise measurement in all three methods only considers the adjacent vertices and some non-adjacent pathways that could possibly have a significant contribution in the robustness of the network, might be neglected.  Here, the subject of interest is more the similarities of the results with respect to the three notions of curvatures that we discussed above.

\subsection{Directed Forman Curvature in \emph{E. Coli} Network}

In the previous sections, we compared three notions of network curvature that apply in the unstructured case, where the network is modeled by an undirected graph. All three notions admit a generalization in which positive numerical edge weights and node weights were taken into account. In this section, we will discuss a further generalization of the Forman-Ricci (FR) curvature in which edge directions and edge ``control types'' are also taken into account.

Real biological networks are almost always directed, in the sense that the edges represent a relationship of action of the agent or quantity labeled by one node upon the agent or quantity labeled by the other. In practice the nature of this action or influence is either positive (promotion, activation) or negative (repression, inhibition). We will use the terms \emph{positive control} and \emph{negative control} for these properties. A network augmented with such edge properties will be called a \emph{signed-control network}. In this section we focus on directed, weighted, signed-control networks.

As already indicated, the authors \cite{Weber} and \cite{Sreejith} imported FR curvature into the domain of network analysis, and demonstrated a certain amount of utility of this tool (recall formula \ref{Forman-curvature-weighted}).
A proposed generalization to the directed case appears in \cite{Sreejith}, consisting of a simple restriction on the terms appearing in \ref{Forman-curvature-weighted}. Namely, for a given edge $e$, one calculates the ordinary FR curvature of the ``source-free and drain-free" subnetwork defined near $e$. By definition this subnetwork contains only those directed edges terminating at the source of $e$ or emanating from the target of $e$. The authors call these edges the edges ``consistent with $e$''.

We reiterate that we assume that the networks which we consider are positively weighted. Each edge weight is interpreted as a quantification of the strength or amplitude of the action represented by the edge. Though mapping directed, positively-weighted, signed-control networks onto directed, signed-weighted networks by signing each edge weight with the control type may have some utility, we think a different approach is more natural.

\begin{figure}[!ht]
\centering
\includegraphics[width=0.4\textwidth]{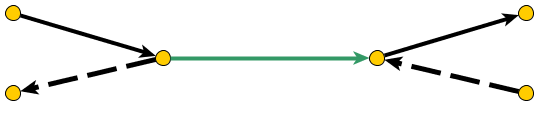}
\caption{The green edge is the edge $e$ under consideration. The dashed edges do not contribute to the directed FR curvature at $e$.}
\label{Directed1Neighborhood}
\end{figure}

Figure~\ref{Directed1Neighborhood} illustrates the classification of the edges in the first-neighbor neighborhood of a given edge $e$ which leads to the modification of formula \ref{Forman-curvature-weighted} for directed networks. The dashed arrows indicate edges whose corresponding terms are omitted.

\begin{figure}[!ht]
\centering
\includegraphics[width=0.4\textwidth]{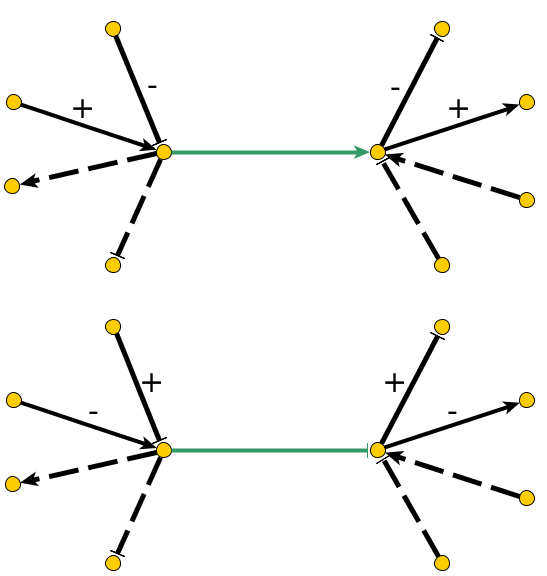}
\caption{As in figure \ref{Directed1Neighborhood}, the dashed edges do not contribute terms to the curvature at green edge $e$. In the respective cases, $e$ of positive and negative control, the contribution of each neighbor is signed as indicated.}
\label{DirectedControl1Neighborhood}
\end{figure}

Figure~\ref{DirectedControl1Neighborhood} shows the analogous classification in the case of directed signed-control networks. Positive control is indicated with an ordinary arrow, while negative control is indicated by a flat arrowhead (this symbol is consistent with usage in \cite{Alon}).

We seek a further modification which takes account of the control types for each edge. We choose to introduce signs to the terms according to the following heuristic, closely resembling the heuristic of Sreejith \emph{et al}  \cite{Sreejith} in establishing the directed generalization. As before, each term corresponds to a certain edge adjacent to $e$, and we only consider the edges consistent with $e$ in the directed sense. In case $e$ and the adjacent edge are both of positive control, the total effect of the pair is considered to be of positive control, and we proceed as in the case or ordinary directed networks, making no alteration to the corresponding term. In case exactly one of $e$ and the adjacent edge is of negative control, the total effect of the pair is considered to be of negative control, and in this case we multiply the corresponding term by the sign $-1$. In case both $e$ and the adjacent edge are of negative control, the total effect of the pair is considered to be of positive control, and in this case no alteration is made to the corresponding term.

We call the resulting curvature notion \emph{signed-control curvature} as an abbreviation for \emph{directed, signed-control Forman-Ricci curvature}, in which the presence of directions and the relation to Forman-Ricci curvature are understood. Also, the scalar signed-control curvature of a given node in the network is defined as the sum of the curvatures of the adjacent edges to that node as before \ref{scalar_FR}.

Notice that the resulting formula is not identical with the formula one would obtain by mapping the original directed, weighted, signed-control network onto a directed, signed-weighted network and calculating the FR curvature. In that case the edges of negative control would always appear with a negative sign, whereas in our case this sign depends also on the control type of the edge $e$ under consideration.

The readers familiar with category theory may wish to conceptualize our construction more succinctly by observing that our signed-control networks can be regarded as generating freely a mathematical category together with a functor to the groupoid $\{\pm 1\}$ (and another functor recording the edge weights). Then the curvature of an edge (morphism) is a certain sum over composable neighboring morphisms, signed by the values of the functor on the composition. This point of view is convenient but not essential.

\begin{figure}
\centering
\includegraphics[width=0.75\textwidth]{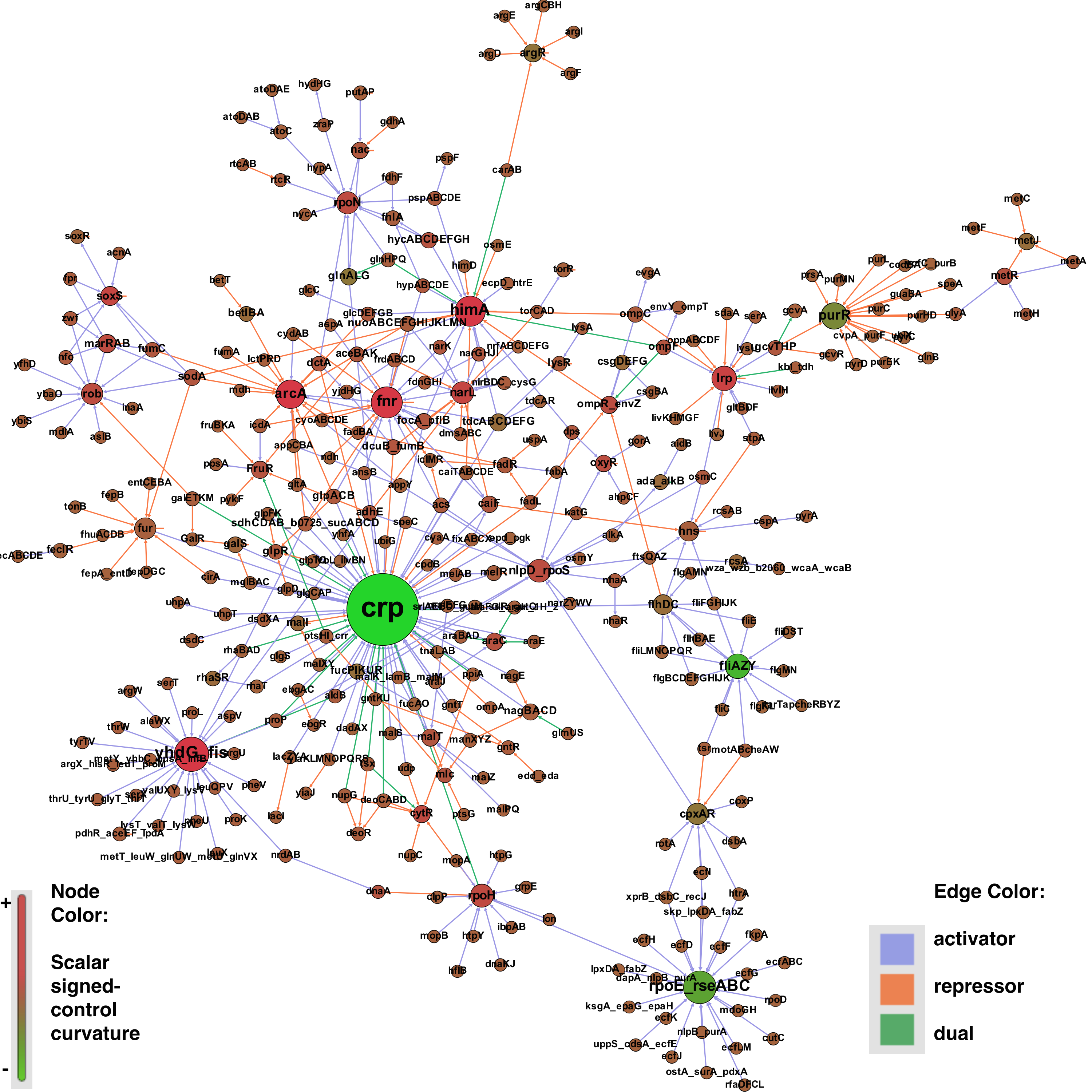}
\caption{The largest connected component of the \emph{E. Coli} transcription network (via Gephi).  The network consists of 423 nodes and 578 edges.  The edges are: 58\% activator, 37\% repressor and 5\% dual (both activator and repressor). The dual edges are considered as activators in the calculation of signed-control curvature.  Here, the size of the nodes are proportional to the node-degree and the colors of the nodes correspond to the scalar signed-control curvature of that node. Also, the color of the edges represents the sign of the regulation. }
\label{figure:EColi}
\end{figure}

We apply the signed-control curvature to the directed signed-control network of \emph{E. Coli} gene products and their transcription regulation relations, obtained from the  Alon website \cite{E.Coli_Alon}.  Table~\ref{table:E.Coli} provides average FR curvature on undirected, directed, and directed signed control forms of \emph{E. Coli} network. Moreover, these notions are provided for five operons of the network with highest degree (sum of In-degree and out-degree).  A visualization of the largest connected component of the network is presented in Figure~\ref{figure:EColi}. As it is shown in the table considering the directions and the signs associated to these directions can have a great impact on the curvature values.

\begin{table}
\caption{Forman-Ricci curvature has been calculated on E. Coli transcription network in three settings: FR curvature on the undirected network, directed FR curvature that consider the directions of the edges, and signed-control FR curvature on the directed and signed-control network. In addition to the average curvature of the E. Coli network, we provide the curvature of five hubs in the network. The degree (sum of in-degree and out-degree) of these nodes has been included in parentheses following the name of the operons.}
\centering
 \begin{tabular}{ |p{2.6cm}||p{1.2cm}|p{1cm}|p{1cm}|}
 \hline
 \bf{ Operon (degree)} & \bf {Undirected FR curvature } & \bf {Directed FR curvature } & \bf {Signed-control (FR) curvature }\\ [0.5ex] 
 \hline\hline
Average curvature of network  & -11.713  & -1.008  & 0.967 \\
\hline
crp (74) & -366 & -214 & -110\\
\hline
yhdG-fis (28) & -499 & -58 & 154\\
\hline
rpoE-rseABC (26) & -713  & -62 & -62\\
\hline
fnr (24) & -26 & -72 & 88\\
\hline
himA (23) & -3 & -56 & 68\\
 \hline
\end{tabular}
\label{table:E.Coli}
\end{table}

\subsection{Comparing Ricci curvature to common network measures}

We are also interested in the relationship between Ricci curvature and common measures of a network.  We study the correlation between Ricci curvature and two important network measures, namely, betweenness centrality and clustering coefficient.

Betweenness centrality (of a node) quantifies the number of times a node acts as a bridge along the geodesic path between two other nodes. More precisely, \emph{betweenness centrality} of a vertex $v$ is defined as \cite{between}:
$$ \text{BC}_v = \sum_{s\neq v \neq t \in V} \frac{\delta_{st}(v)}{\delta_{st}}$$
where $\delta_{st}$ is the total number of geodesic paths from node $s$ to node $t$ and $\delta _{st}(v)$ is the number of those paths that pass through $v$.

\begin{figure*}[ht!]
     \begin{center}
        \subfigure[]{%
           \label{}
           \includegraphics[width=0.9\textwidth]{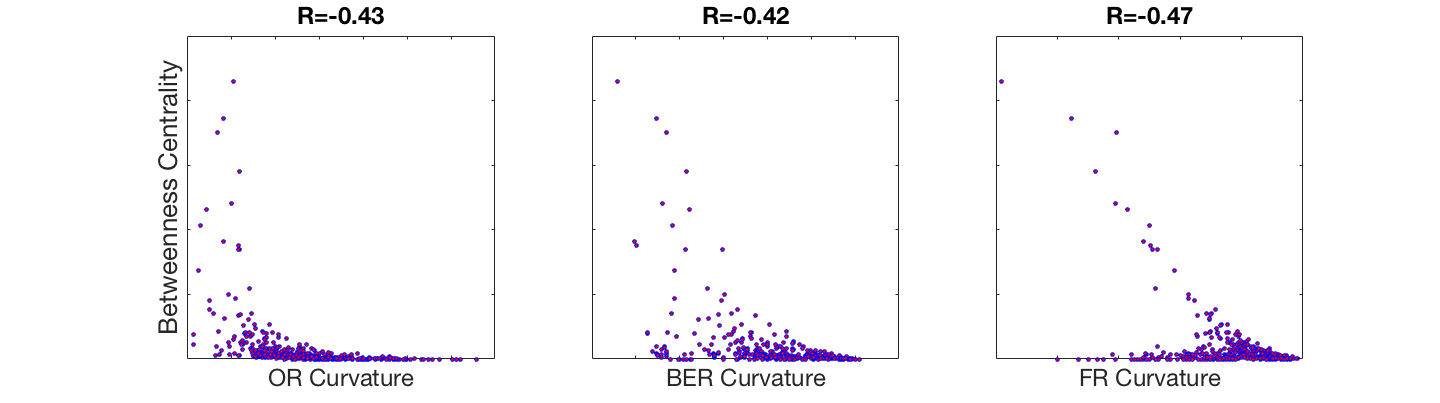}
        }%
        \\
        \subfigure[]{%
            \label{}
            \includegraphics[width=0.9\textwidth]{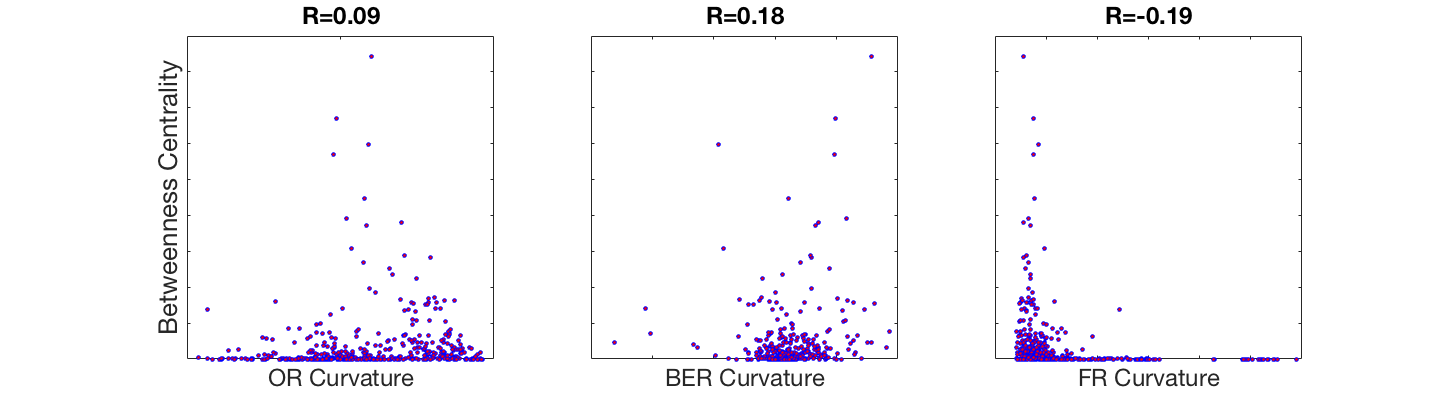}
        }%

    \end{center}
    \caption{(a) Scale-free Network (b) Breast Cancer Network, (Pearson) correlation between three notions of curvature and betweenness centrality for each node of the network. The shapes of the correlation and the values of the correlation coefficient (R) are similar with these three different notions of discrete Ricci curvature.}
   \label{fig:3}
\end{figure*}

Figure~\ref{fig:3} shows a correlation between three notions of curvature and betweenness centrality in a scale free network model and the cancer network of breast carcinoma.  The correlation is slightly stronger in the scale free network but overall there is weak correlation between these two measured on the network which suggests that they are measuring different assets in the network.  It is also interesting to note that the overall shape of the correlation plot and correlation coefficient value is very similar with different discrete notions of Ricci curvature for the two networks.  This can also help us to understand that these curvatures are presenting the same measure on the complex networks.

On the other hand, a clustering coefficient is a measure of the relative local frequency of triangles within a graph. The local clustering coefficient of a vertex $v$ in graph is introduced by Watts-Strogatz \cite{Cluster} as follows:
$$ C_v= \frac {\text{number of triangles connected to node} \, v}{\text{number of triples centered around node} \,  v}$$
where a triple centered around node $v$ is a set of two edges connected to node $v$ (If the degree of node $v$ is 0 or 1, which gives us $C_v=0/0$, we set $C_v=0$). By definition, $0\leq C_v \leq 1$.

\begin{figure*}[ht!]
     \begin{center}
        \subfigure[]{%
           \label{}
           \includegraphics[width=0.9\textwidth]{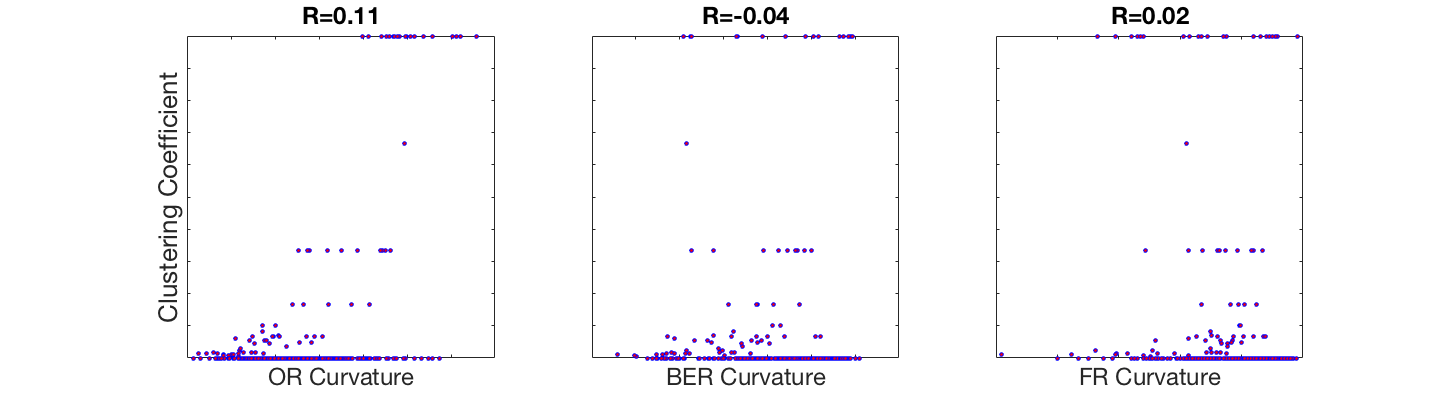}
        }%
        \\
        \subfigure[]{%
            \label{}
            \includegraphics[width=0.9\textwidth]{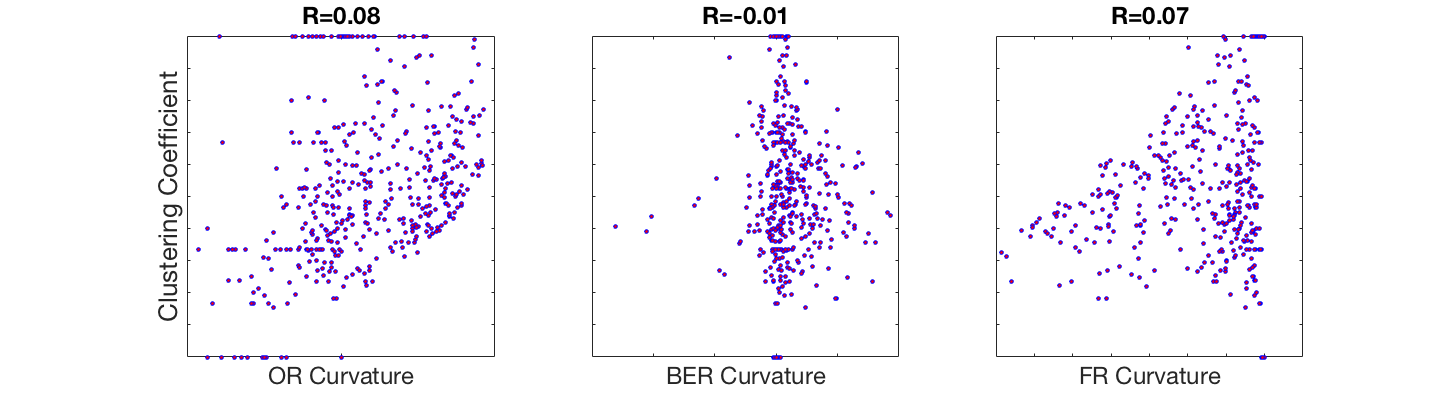}
        }%

    \end{center}
    \caption{(a) Scale-free Network (b) Breast Cancer Network, (Pearson) correlation between three notions of curvature and clustering coefficient for each node of the network. The shapes of the correlation and the values of the correlation coefficient (R) are similar with these three different notions of discrete Ricci curvature.  }
   \label{fig:4}
\end{figure*}

As discussed by Jost and Liu \cite{jost}, local clustering coefficients can define a lower boundary for the Ollivier-Ricci curvature. In fact, they discuss that a lower Ricci curvature bound prevents geodesics from diverging too fast and balls from growing too fast in volume. An analogue of geodesics starting in different directions, but eventually approaching each other, could be a triangle on a graph. Therefore, they expect that the Ricci curvature on a graph be related to the relative abundance of triangles. However, in practice the correlation between clustering coefficient and Ricci curvature on a graph is not very strong.  This can be seen in Fig. \ref{fig:4}.  Again, this suggests that these two are in fact measuring the network differently.  Also, the plots for each network with three different Ricci curvature look pretty similar.

\section{Discussion}

We believe that Ricci curvature can help us to study the robustness of complex networks. Consequently, we can quantify the differences between normal and robust networks by computing their Ricci curvature.   We reviewed three techniques that can be used for measuring the Ricci curvature on networks.  Our methods are very much intertwined with the theory of optimal mass transport.  In our findings, all seven studied cancer networks have shown a higher average Ricci curvature than the normal complementary networks for all methods of computation.  There is not a very strong correlation between Ricci curvature and betweenness centrality nor the clustering coefficient.  However, all of the results are consistent with three different notions of discrete Ricci curvature on networks. Therefore, understanding the geometry of the networks through the Ricci curvature yields a novel measurement which is connected to the entropy and robustness of networks.

The first point of interest for future work is to repeat this study on larger networks, using data consisting of both cancer and non-cancer-related genes. Ranking the nodes by their Ricci curvature, we would expect to find the cancer-related genes clustered at the top of the list and the non-cancer-related genes clustered at the bottom. This would provide additional support for our definition of Ricci curvature and any unexpected rankings could possibly provide some new insight.  Furthermore, we are interested in generalizing this work from scalar-valued densities to the vector-valued ones.  We are interested in applying some recent results \cite{Chen1} on ``vector-valued'' mass transport theory to ``multiomics'' cancer networks in which both protein and genomic data are combined rather than treated as independent entities. We also want to derive more computationally efficient methods for computing the 1-Wasserstein metric as a clustering technique that has found some success in a recent work on sarcoma \cite{Chen2}.

In summary, our study shows the consistency of the results using three very different discrete analogues of Ricci curvature on biological cancer networks. Perhaps more importantly, we introduced the notion of \emph{\textbf{signed-control curvature}} on (directed) signed-control networks which can be applied to many real biological networks equipped with positive/ negative controls and directions.

\section*{Acknowledgements}
This project was supported by AFOSR grants (FA9550-15-1-0045 and FA9550-17-1-0435), ARO grant (W911NF-17-1-049), grants from the National Center for Research Resources (P41-RR-013218) and the National Institute of Biomedical Imaging and Bioengineering (P41-EB-015902), NCI grant (1U24CA18092401A1), NIA grant (R01 AG053991), MSK Cancer Center Support Grant/Core Grant (P30 CA008748), and a grant from the Breast Cancer Research Foundation.


\begin{table*}[ht]
\begin{tabular}{ |p{1 cm}||p{2cm}|p{2cm}|p{2cm}|}
\hline \bf{Gene Ranking} & Gene A  & Gene B &  $\Delta$ OR Curvature (Cancer-Normal) \\ [0.5ex]
\hline\hline
1 & RNF43 & RSPO3 & 0.3504 \\
\hline
2 & RNF43 & RSPO2 & 0.3444\\
\hline
3 & ERG & ETV1 & 0.3012\\
\hline
4 & GPC3 & PTCH1 &  0.3001\\
\hline
5 & SDC4 & GPC3 & 0.2901\\
\hline
6 & POT1 & SBDS & 0.2796 \\
\hline
7 & FGFR2 & KDR & 0.2538\\
\hline
8 & ERG & FOXA1 & 0.2460\\
\hline
9 & SDC4 & EXT1 & 0.2410\\
\hline
10 & MYC & SDHD & 0.2408\\
\hline
11 & TAL1 & RUNX1 & 0.2167\\
\hline
12 & SDC4 & EXT2 & 0.2165\\
\hline
13 & NUP214 & ELN & 0.2132\\
\hline
14 & TAL1 & TCF3 & 0.2123\\
\hline
15 & PDGFB & COL2A1 & 0.2036\\
\hline
16 & IDH1 & IDH2 & 0.2012\\
\hline
17 & SDHB & HMGA1 & 0.2007\\
\hline
18 & TAL1 & TRIM27 & 0.1929\\
\hline
19 & EPS15 & MLLT4 & 0.1909\\
\hline
20 & FUBP1 & PICALM & 0.1899\\ [1ex]
\hline
\end{tabular}
\caption {The top 20 pairs of genes with respect to Olivier-Ricci curvature.}
\label{table:8}
\end{table*}

\begin{table*}[ht]
\begin{tabular}{ |p{1 cm}||p{2cm}|p{2cm}||p{1 cm}||p{2cm}|p{2cm}|}
\hline \bf{Gene Ranking} & Gene &  $\Delta$ BER Curvature (Cancer-Normal) & {Gene Ranking} & Gene &  $\Delta$ BER Curvature (Cancer-Normal)\\ [0.5ex]
\hline\hline

1 & PICALM & 7.4910 & 21 & PDGFRB & 2.8796\\
\hline
2 & CLTCL1 & 4.9102 &  22 & JAK2 & 2.7667 \\
\hline
3 & EPS15 & 4.3210 &  23 & RPN1 & 2.6685\\
\hline
4 & KIF5B & 4.1284 & 24 & DCTN &  2.4304\\
\hline
5 & CLTC & 4.0657 & 25 & TBL1XR1 & 2.3959 \\
\hline
6 & PTPN11 & 4.0465 & 26 & ABL1 & 2.3943\\
\hline
7 & YWHAE & 3.8416 & 27 & PIM1 & 2.3879\\
\hline
8 & EGFR & 3.8357 & 28 & PBRM1 & 2.3454 \\
\hline
9 & JAK1 & 3.7590 & 29 & TFRC & 2.3162\\
\hline
10 & MSN & 3.6079 & 30& NDRG1 & 2.2251\\
\hline
11 & CDC73 & 3.5274 & 31& LCK & 2.1788\\
\hline
12 & PIK3CA & 3.4499 & 32 & KIT & 2.1602\\
\hline
13 & XPO1 & 3.4274 & 33 & FGFR1& 2.1367\\
\hline
14 & ALDH2 &  3.3854 & 34 & STAT5B & 2.0506 \\
\hline
15 & SDHB & 3.2626 & 35 & ERG & 2.0441\\
\hline
16 & GNAS & 3.1372 & 36 & KDR & 2.0387 \\
\hline
17 & AKT1 & 3.1279 & 37& PPARG & 2.0034\\
\hline
18 & MAP2K1 & 3.0754 & 38 & SYK & 1.9919\\
\hline
19 & CBL & 3.0287 & 39 & HIP1 & 1.9897 \\
\hline
20 & PML &  3.0043 & 40 & CUX1 & 1.9819\\ [1ex]
\hline
\end{tabular}
\caption {The top 40 genes with respect to local BER curvature.}
\label{table:9}
\end{table*}

\begin{table*}[ht]
\begin{tabular}{ |p{1 cm}||p{2cm}|p{2cm}|p{2cm}|}
\hline \bf{Gene Ranking} & Gene A  & Gene B &  $\Delta$ FR Curvature (Cancer-Normal) \\ [0.5ex]
\hline\hline
1 & CARS & ALDH2 &  256.904\\
\hline
2 & NDRG1 & ARNT & 252.955\\
\hline
3 & ALDH2 & SDHB & 239.857\\
\hline
4 & CLTCL1 & ALDH2 &  227.896\\
\hline
5 & ALDH2 & KIF5B & 219.595\\
\hline
6 & EBF1 & CEBPA & 218.589 \\
\hline
7 & CLTCL1 & SDHB & 212.586 \\
\hline
8 & CEBPA & PPARG & 209.869 \\
\hline
9 & PTPN11 & PTPRB & 208.132 \\
\hline
10 & ALDH2 & IDH1 & 198.918 \\
\hline
11 & EPS15 & SDHB & 195.605\\
\hline
12 & CDH1 & FUS & 193.463\\
\hline
13 &  SDHB & PTPN11 & 191.581\\
\hline
14 & JAK1 & AKT2 & 191.086\\
\hline
15 & ELF4 & ERG & 188.118\\
\hline
16 & AKT2 & PIK3CA & 187.711\\
\hline
17 & NONO & RUNX1 & 186.994\\
\hline
18 & HMGA1 & RARA & 186.925\\
\hline
19 & CLTCL1 & HIP1  & 186.329\\
\hline
20 & ERBB2 & ATP1A1 & 185.469\\ [1ex]
\hline
\end{tabular}
\caption {The top 20 pairs of genes with respect to FR curvature.}
\label{table:7}
\end{table*}

\end{document}